 \documentclass{article}

\usepackage{graphicx}
\usepackage{dcolumn}
\usepackage{bm}
\usepackage{wrapfig}
\usepackage{caption}
\usepackage{amsmath}
\usepackage{subcaption}
\usepackage{geometry}
\usepackage[colorlinks=true, linkcolor=blue, citecolor=blue, urlcolor=blue]{hyperref} 
\usepackage{units}
\usepackage{float}
\usepackage{orcidlink}
\usepackage{hyperref}

\newcommand{\Lint}{\ensuremath{\mathcal{L}_\mathrm{int.}}}

\usepackage[mathlines]{lineno}

\begin{document}


\title{Energy and Time Histograms:\\Fluxes determination in calorimeters}

\author{Khalid Hassouna\thanks{University of Hawaii at Manoa,
2500 Campus Rd., HI, USA} \orcidlink{0009-0004-8537-9490}
\and Vincent Boudry\thanks{ Laboratoire Leprince-Ringuet (LLR), CNRS, École polytechnique, Institut Polytechnique de Paris, 91120 Palaiseau, France.} \orcidlink{0000-0002-1258-1338}}

\date{March 08, 2024}

\maketitle

\begin{abstract}
In high-granularity calorimetry, as proposed for detectors at future Higgs factories, the requirements on electronics can have a strong impact on the design of the detector, especially via the cooling and acquisition systems. 
This project aims to establish the typical fluxes in the calorimeters: deposited energy, number of cells above the electronics threshold, additional heat, and associated data, etc. Here, a software package is presented which outputs histograms for different selections of the calorimeter components from fully simulated physics and background events. The ILD calorimeter system is taken as a specific example, upon which different histograms are obtained for representative parts of the calorimeter and for various machine configurations. Examples of histograms are shown, along with all details of the data used and the simulation.
\end{abstract}

\section{Introduction \label{sec:intro}}

One of the ways to improve the performance of the detectors for future Higgs, Electroweak or top factories (such as the ILC, CLIC, CEPC and FCC-ee) is to make use of the Particle Flow Algorithms~\cite{brient_calorimetry_2002}; this is the approach adopted for e.g.\@ the SiD~\cite{breidenbach_updating_2021}, ILD~\cite{the_ild_collaboration_international_2020}, CLICdet~\cite{linssen_physics_2012}, CEPC baseline~\cite{the_cepc_study_group_cepc_2018} and CLD~\cite{bacchetta_cld_2019} detector concepts. The PFA performance relies on imaging calorimeters, able to distinguish the individual contributions from neutral and charged particle showers, associated with a precision tracker system. 
The required granularity is typically below the centimetre scale for the ECAL, and slightly above for the HCAL~\cite{thomson_particle_2009, the_cepc_study_group_cepc_2018}, resulting in 10 to 100 million channels for each of them. \\

The original designs, SiD and ILD, have been made for the ILC running conditions, with center-of-mass (CM) energy of 250 to 1000\u{GeV}, and bunch rate of 5\u{Hz}. Since then, other colliders have been proposed with widely different running conditions: CLIC can reach a much higher CM energy (3\u{TeV}) and repeating rates.  FCC-ee and CEPC would have continuous operation with huge luminosity for the Z-boson peak at a CM energy of 91.2\u{GeV}. \\

Given the need to potentially adapt the calorimetric system to various conditions, estimating the static and dynamic behaviour of particle energy deposition became necessary. The former can be represented by the energy distribution along with its spatial derivatives (radius, azimuthal angle), and the latter can be represented by the time distribution along with its derivatives. These distributions will serve as a guide to the power needed from the electronics to optimize the electronics and prevent them from overshooting the target, thereby ensuring they do not adversely affect the calorimeter's physics. \\

This paper presents a software package \footnote{It can be found here: \url{https://github.com/LLR-ILD/CalorimeterFluxes}} that enables obtaining the desired distributions for any chosen part of the calorimeter from a full simulation of the main expected physics processes.  We apply it here first on the simulated ILD calorimeter, which is a very suitable choice because the ILD calorimeter is divided into logical coordinates, which facilitates the selection of various parts of the calorimeter. \\

The paper is divided into four parts, starting with an introduction that presents the terminology used, in detail. In the next part, the produced sets of histograms are explained, including their description and classification. The used data sample is then detailed, and finally, we present some results along with preliminary conclusions. 

\section{Terminology}
\label{sec:terminology}

The calorimetric system of ILD is composed of 11 systems; some are redundant as options.
Two options are proposed (and simultaneously simulated) for the ECAL, named after their sensor's technology: ScECAL for the scintillators and SiW-ECAL for the Silicon.
Also, two options are available for the HCAL. The AHCAL is based on tiles of scintillators, while the SDHCAL's sensors are segmented Resistive Plate Chambers.
Both the ECAL and HCAL have two main parts: the Barrel for the central region and the Endcaps for the forward and backward regions. The Barrel has an inner shape of an octagonal prism, closed by 8 "Staves". The Endcaps inner shape is a square, formed by 4 quadrants.
They are complemented by “rings” filling the gaps between the Barrel and Endcaps for the HCAL, and between the Endcaps and the beam tube for the ECAL.
Each system may be divided into a hierarchy of sub-regions: stave or quadrant, modules, towers, layers, cells, each addressed by a logical coordinate: C:S:M:T:L:I:J\footnote{C stands here for the calorimeter System, S for the Stave, M for the module, T for the tower, L for the layer, I and J for the logical coordinates}. This segmentation follows that foreseen for the electronics, and therefore might be a bit complex in terms of geometry. \\

A hit is defined as a cell whose energy exceeds the electronics threshold defined for the system. The threshold is set at 1/3rd or 1/4th of a "MIP" energy scale, defined as the most probable value of the energy distribution yielded by a minimum ionizing particle (in practice medium energy muons) crossing the sensor with a normal incidence.

\subsection{Direct and Indirect Selections}
\label{sub:selections}

In our analysis, we distinguish between what we call direct selections and indirect selections. The former is given by directly stating the calorimeter coordinates. For example, the selection \{system = ScEcalBarrel, Stave = 1, Module = 2, Tower = 5\} is a direct selection of the calorimeter coordinates. An indirect selection will be given through what we call Boolean complex functions. For instance, the selection \{all modules and layers satisfying the equation: $2\mathrm{Module} + 2\mathrm{Layer} = 10$\} is an indirect selection because the choices of modules and layers depend on each other through an equation (the same as specifying a boolean function that is true if and only if the coordinates satisfy the given equation). If no selection is imposed on a coordinate, all such coordinates will be included. For instance, in the direct selection example above, no condition is requested on the layers; this means that all the layers are included. The same applies to the indirect selections. \\

The introduction of these indirect selections was necessary in some special cases to restore the symmetry in some distributions as the direct selections were not enough in providing a separation of distributions, as will be shown later.  

\subsection{Primary and Secondary histograms}
\label{sub:histograms}

The histograms in which we store direct information from the hits and events are called primary, while the histograms that are built from other histograms are called secondary. For example, one important histogram is that of energy. The primary histogram would have the unit of the X-axis to be in GeV. On the other hand, one secondary histogram would be exactly the same, but with the unit being a MIP. This secondary histogram is just a scaling of the X-axis of its primary counterpart. 

\subsection{1D and 2D histograms}
\label{sub:dimensional histograms}

\begin{figure}[htbp]
    \centering

    \begin{subfigure}[b]{0.3\textwidth}
        \includegraphics[width=\linewidth]{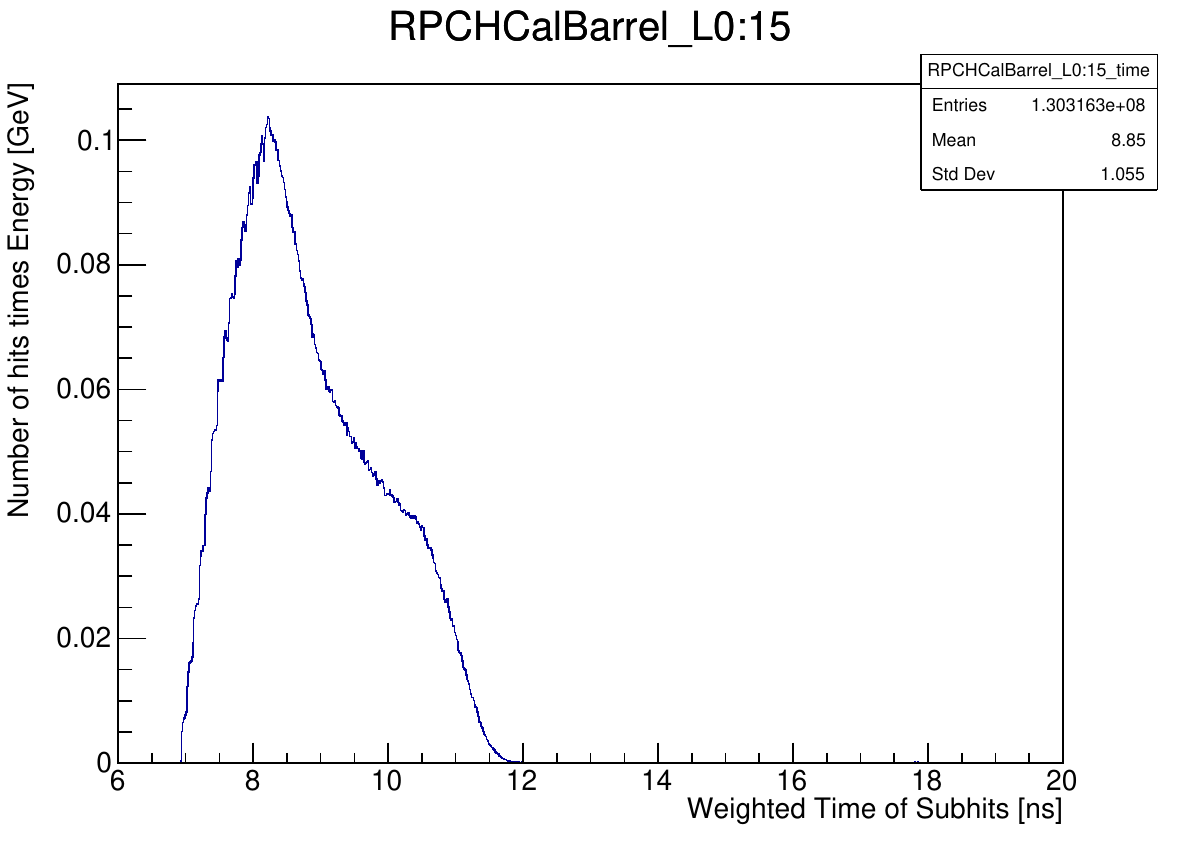}
        \caption*{Energy-weighted time distribution of the first 16 layers}
    \end{subfigure}%
    \hspace{10pt}
    \begin{subfigure}[b]{0.3\textwidth}
        \includegraphics[width=\linewidth]{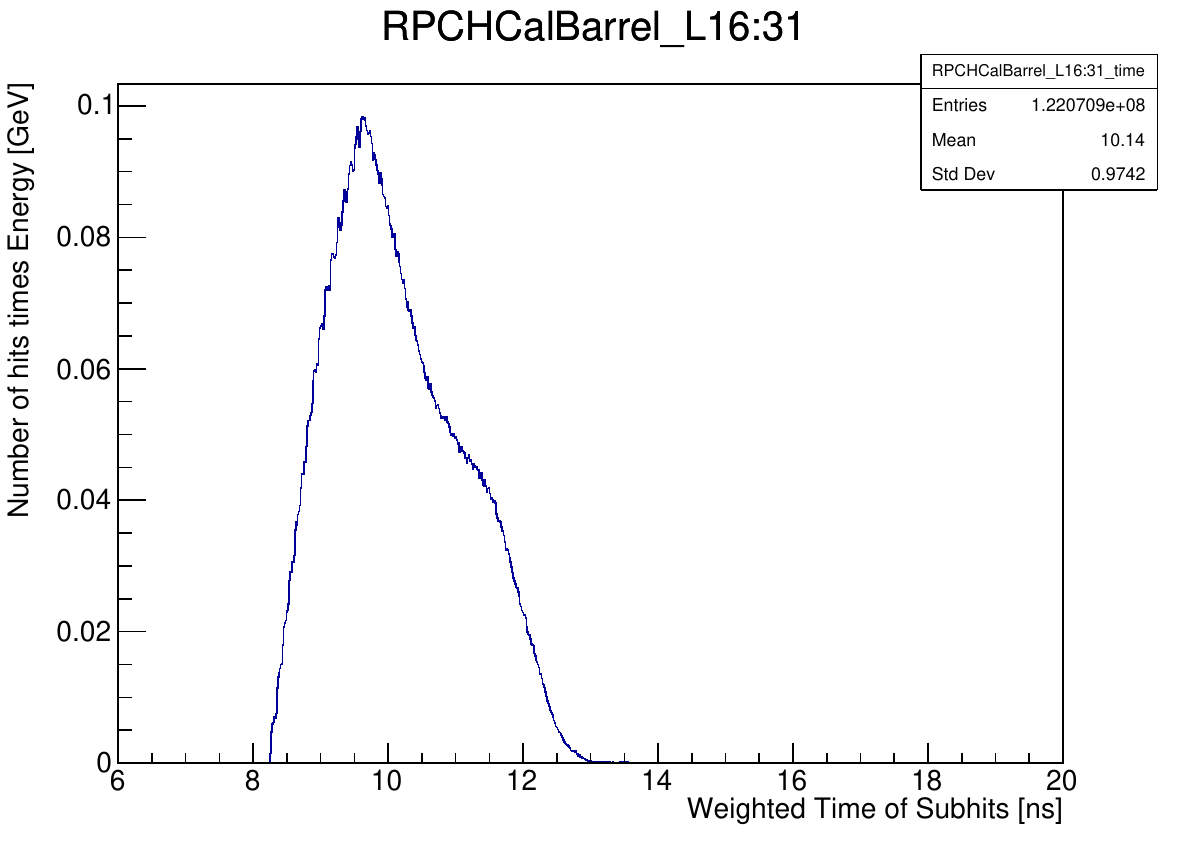}
        \caption*{Energy-weighted time distribution of the second 16 layers}
    \end{subfigure}%
    \hspace{10pt}
    \begin{subfigure}[b]{0.3\textwidth}
        \includegraphics[width=\linewidth]{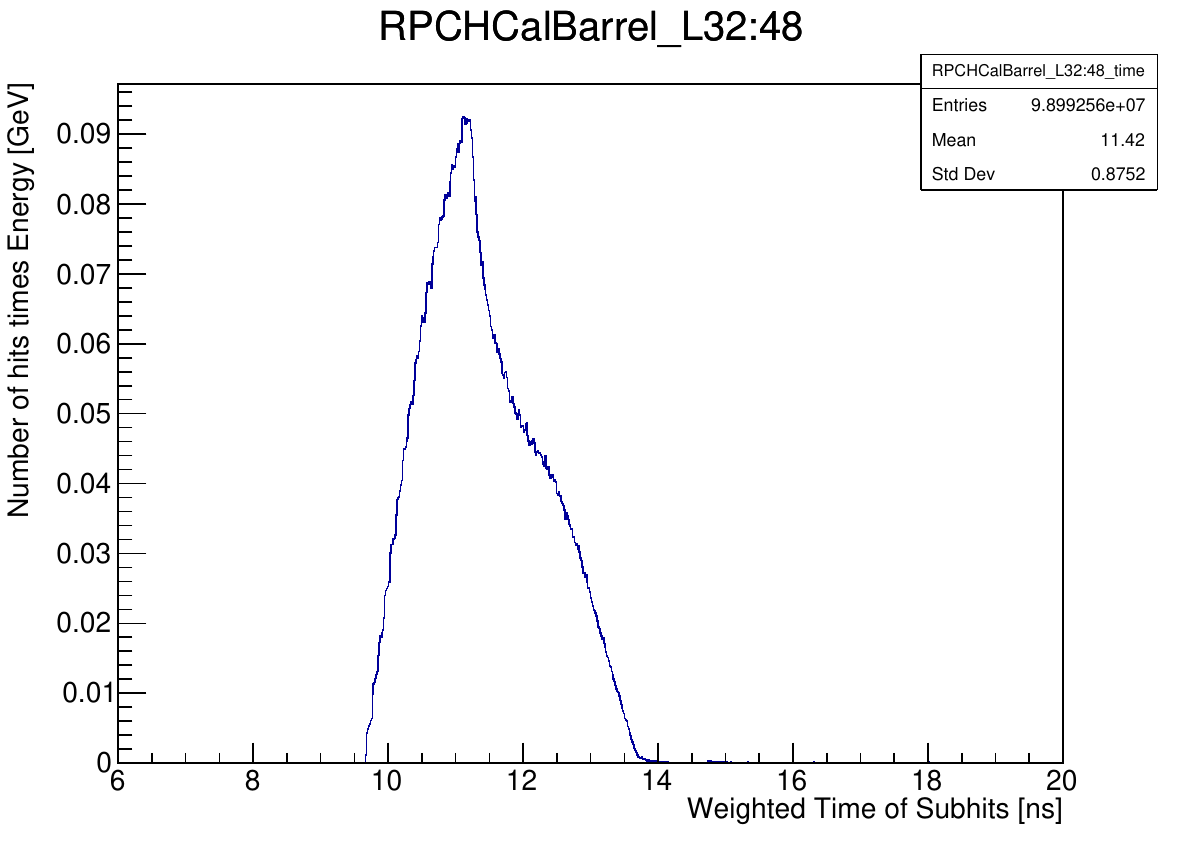}
        \caption*{Energy-weighted time distribution of the third 16 layers}
    \end{subfigure}%

    \begin{subfigure}[b]{0.9\textwidth}
        \includegraphics[width=\linewidth]{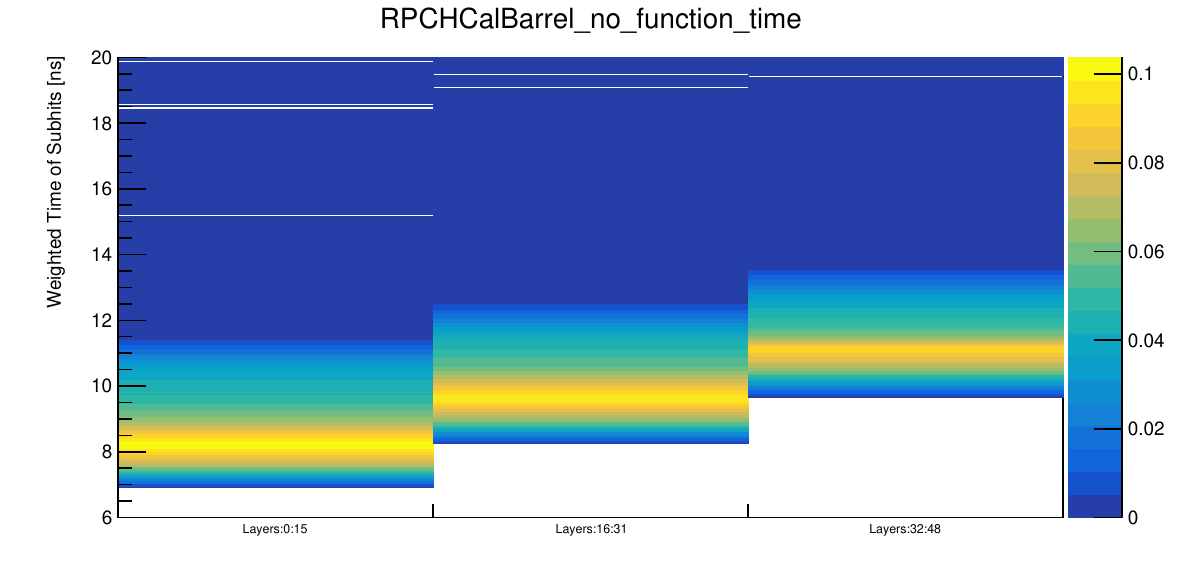}
        \caption*{Energy-weighted time distribution of all the layers in RPCHCalBarrel system, where each bin represents a 1D histogram. `no\_function' in the title stands for direct selections (no boolean functions involved for indirect selections)}
    \end{subfigure}

    \caption{Time distributions of shooting muons in the RPCHCalBarrel system, for verification purposes. The three 1D histograms represent the energy-weighted time distribution for three sequential groups of 16 layers each. The 2D histogram is the collection of the 1D histograms in one plot. The 2D histograms can be very helpful in visualizing the angular distribution. It can be seen clearly that there is a time shift because the muons travel in the layers in order.}
    \label{fig:1D Vs. 2D histograms}
\end{figure}

For practical reasons, the primary histograms are 1D, meaning the Y-axis represents the number of hits, events, or energy-weighted hits, while the X-axis corresponds to a physics quantity such as energy, time, or number of hits. To enhance clarity, multiple 1D histograms are combined into a single 2D histogram, where each bin represents a 1D histogram. For example, consider the energy distribution of the following direct selection: \{System = ScEcalBarrel, Module = 1, Layer=(0-9)\}. This selection will give the energy distribution for the system ScEcalBarrel in its module 1 and the first 10 layers (and indirectly for all staves and cells). Considering the same distribution, but with the layers being from 10 to 19 and another one with the layers ranging between 20 and 29, will produce 3 different 1D histograms. We can put them together in one 2D histogram with 3 bins, such that each 1D histogram represents a single bin. The 1D Y-axis becomes the 2D Z-axis, while the 1D X-axis transitions to the 2D Y-axis. This will give an aid in visualization and will show the angular distribution. An illustrative example is shown in figure \ref{fig:1D Vs. 2D histograms}.  

\section{Histogram Sets}
\label{sec:histograms sets}

As indicated in the introduction, the software is versatile in that it can be easily adapted to produce any types of desired histograms. For our current analysis, we are concerned with mainly 8 histograms: 5 of primary type and 3 of secondary type.

\subsection{Primary Histograms}
\label{sub:primary histograms}

Some distributions are split into low-scale and high-scale parts; the low-scale part uses fixed, high-density binning, while the high-scale part starts at the end of the low-scale and is automatically re-scaled upwards to capture all the tails. The binning is defined so that a complete spectrum can be rebuilt by regrouping bins from the low-scale part.

The following primary histograms have been defined:
\begin{enumerate}
    \item \textbf{Lower-Scale Energy:} This is the histogram of the energy distribution of simulated calorimeter cells. It can be used to define the MIP of a system, and check the incidence of the thresholds variations on the number of hits. Depending on the type of sensor,  it will peak near 0 or at MIP scale.
    \item \textbf{Upper-Scale Energy:} This histogram capture the high energy tails of the cells. After calibration, it will give access to the dynamic range required for the electronics.
    \item \textbf{Lower-scale Number of Hits}: This histogram stores the number of hits until a certain value. This aids in identifying which events contribute most to data flux and the power consumption from conversion (ADC and TDC), under various hypotheses of noise, and electronics treatments, emulated by the digitization. The X-axis is the number of hits and the Y-axis is the number of events. 
    \item \textbf{Upper-Scale Number of Hits:} This histogram captures the high number of hits tails. Combined with the lower-scale part, it allows calculating the average number of hits per event, hence the data rates when scaled by the luminosity and cross-section.
    \item \textbf{Weighted Time:} This histogram presents the time distribution of simulated calorimeter subhits, with each weighted by its energy. This will end up with the energy plotted as a function of time. This histogram give the overall timing behaviour, which is useful to design the electronics making use of the timing in the particle showers reconstruction. \\
    Currently, this corresponds to the lower-scale part (up to 20 ns). A higher-scale part would be useful to estimate the contribution from neutrons of previous events, and will be implemented.
\end{enumerate}

\subsection{Secondary Histograms}
\label{sub:secondary histograms}

\begin{enumerate}
    \item \textbf{Scaled Upper-Scale Energy:} This is just scaling the Upper-Scale-Energy X-axis' histogram by dividing it by the MIP. It is helpful to see the trailing effect as a function of the MIP instead of the normal GeV unit that is not always illuminating.
    The maximum range of this histogram give the dynamic range of the electronics, when divided by the precision needed on the low-energy scale (i.e.\@ the precision on the MIP).
    \item \textbf{Lower-Scale Power:} The power histogram is always a function of the primary histogram, but the exact function form depends on the electronics implementation. Usually, it is one of two things: either expressed through the following equation: $Power = a\times Energy + b$ where $a$ and $b$ are real positive numbers with $a$ representing the proportionality constant between the power and energy and $b$ is the power-on offset of the electronics or $Power = a\times \#\text{hits-above-given threshold}$. The latter case is for when ADCs are only triggered at some energy with a constant power consumption.  
    \item \textbf{Upper-Scale Power:} This is just applying the power equation to the upper-scale-energy histogram to cover all the power required from the electronics. 
\end{enumerate}

\section{Geometric Selections \label{sec:selections}}

This section features geometric plots generated by shooting medium-energy muons. These plots display the hits' geometric coordinates on the x-axis and y-axis, while a colour bar on the z-axis represents properties such as staves, modules, etc. 

\subsection{Direct Geometric Selections \label{subsec:direct}} 
\begin{wrapfigure}[16]{r}{0.40\textwidth}
    \includegraphics[width=6cm, height=5cm]{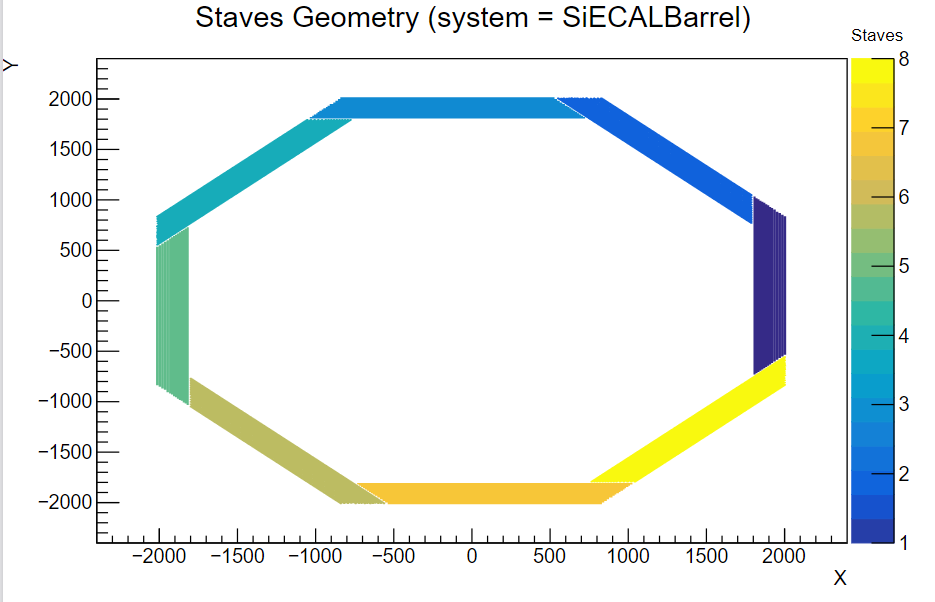}
    \caption{Staves numbering of cells of the SiECalBarrel system in the $x \text{-} y$ plane}
    \label{fig:ECalBarrel Staves Geometry}
\end{wrapfigure}

Geometric selections should be made such that different selections would be grouping obvious symmetric regions, while yielding distributions illustrating the evolution of asymmetric ones. Taking the Silicon-ECAL-Barrel system as an illustrative example, the figure \ref{fig:ECalBarrel Staves Geometry} shows how the staves are distributed in an azimuthal symmetric manner as the beam originates from the origin. Thus, there is no need to make a selection on a specific stave as they all yield the same distributions. Additionally, because there are 30 layers, it is impractical to select each layer individually; doing so would result in many similar distributions, which is not useful. The categorization here is that the layers are always categorized into three blocks with each block containing 10 layers. \\

\begin{figure}[hbtp]
    \centering
    \begin{subfigure}{0.40\textwidth}
    \centering  
    \includegraphics[width=5cm, height=5cm]{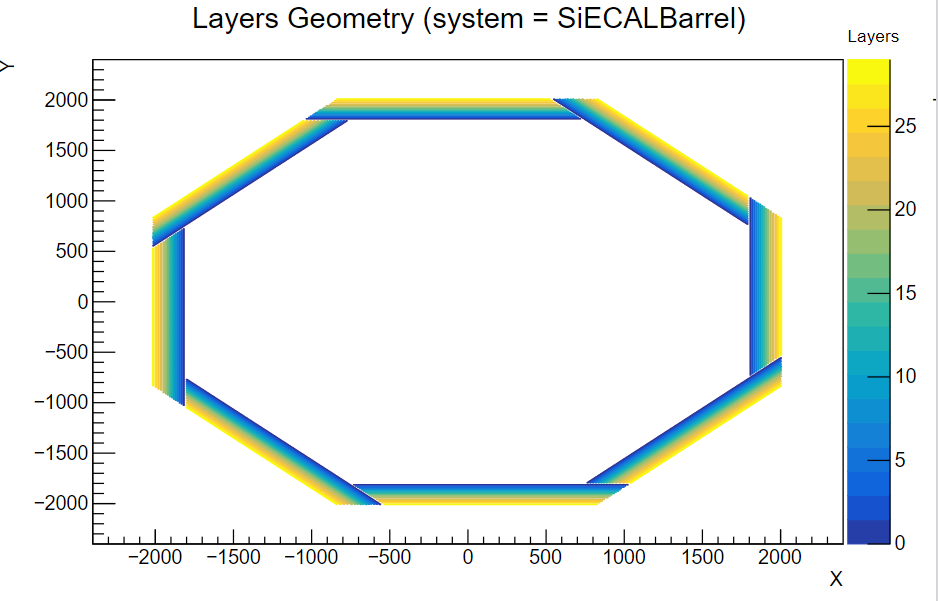}
        \caption{Layers geometry}
        \label{subfig:ECAL layers}
    
    \end{subfigure}%
    \hspace{10pt}
    \begin{subfigure}{0.55\textwidth}
           
        \includegraphics[width=\linewidth, height=5cm]{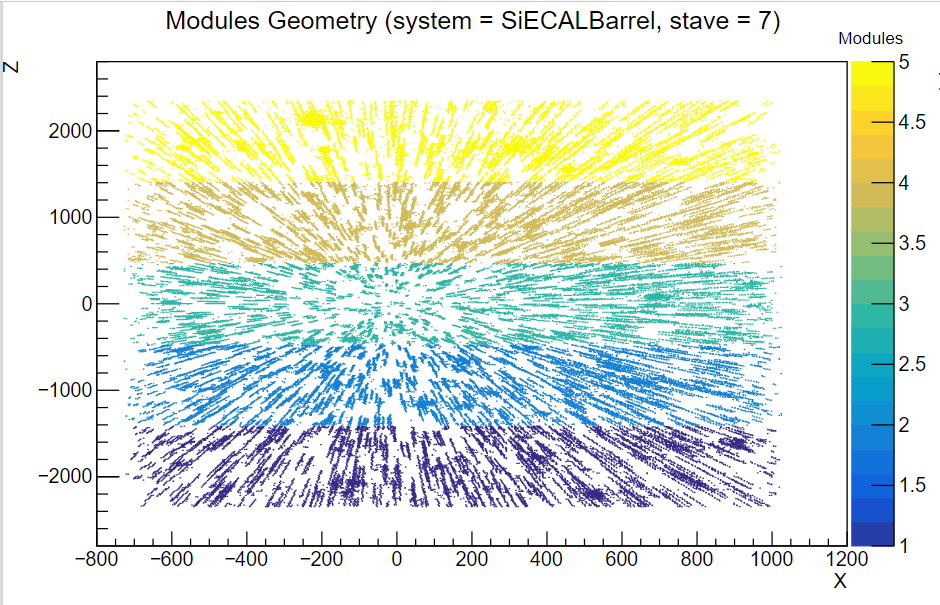}
        \caption{Modules geometry}
        \label{subfig:ECAL modules}
    
    \end{subfigure}%
    \caption{Layers and modules geometry of the Silicon-ECAL-Barrel system}
    \label{fig:ECalBarrel layers-modules geometry}   
\end{figure}

Referring to figure~\ref{subfig:ECAL layers}, it is clear that the layers' number is an indicator of the depth or the radial profile. To get the polar-angle profile, the extension of one of the staves (arbitrarily chosen as all the staves yield the same behaviour) is shown along the Z-axis in figure~\ref{subfig:ECAL modules}. The central module (module 3) is the one at $\theta = 90^{\circ}$ and the other modules present deviations from this right angle. To study the evolution of the distributions along the polar angle, different selections must be made on different modules. It is true that module 2 and 4 are symmetric, and the same applies to modules 1 and 5; however, making a selection on each of them would serve as a cross-check with the prediction that the yielded distributions are to be symmetric. In conclusion, to study the radial profile, three selections are made on the layers, with each selection corresponding to one block of 10 layers. To study the polar-angle profile, five selections are made on the modules, with each selection corresponding to one module. In total, this approach results in $3 \times 5 = 15$ selections.  

\subsection{Indirect Geometric Selections}
\label{subsec:indirect} 

\begin{figure}[H]
    \centering
    \begin{subfigure}{0.45\textwidth}
        \centering
        \includegraphics[width=5cm, height=5cm]{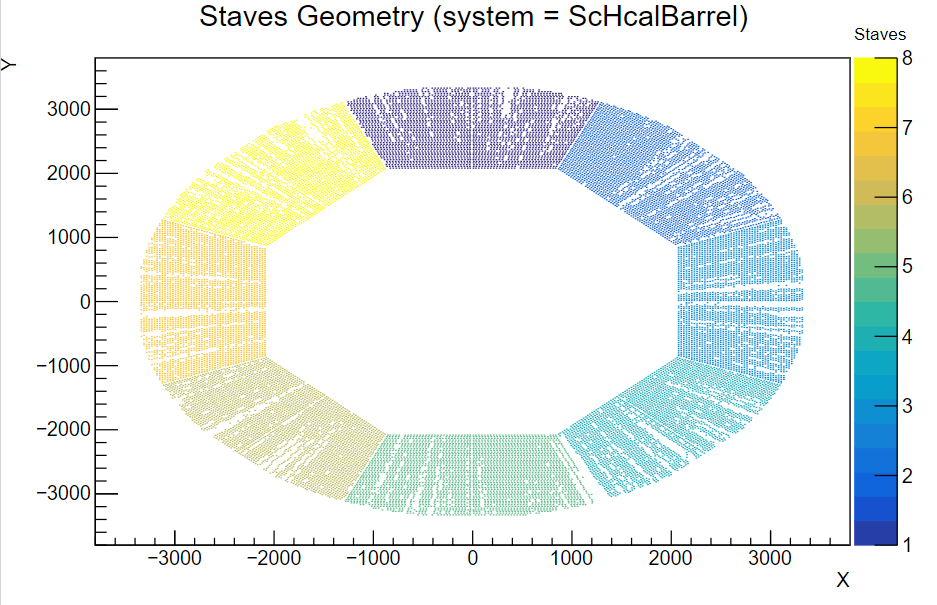}
        \caption{Staves geometry}
        \label{subfig:HCAL staves}
    \end{subfigure}%
    \hspace{10pt}
    \begin{subfigure}{0.45\textwidth}
        \centering
        \includegraphics[width=5cm, height=5cm]{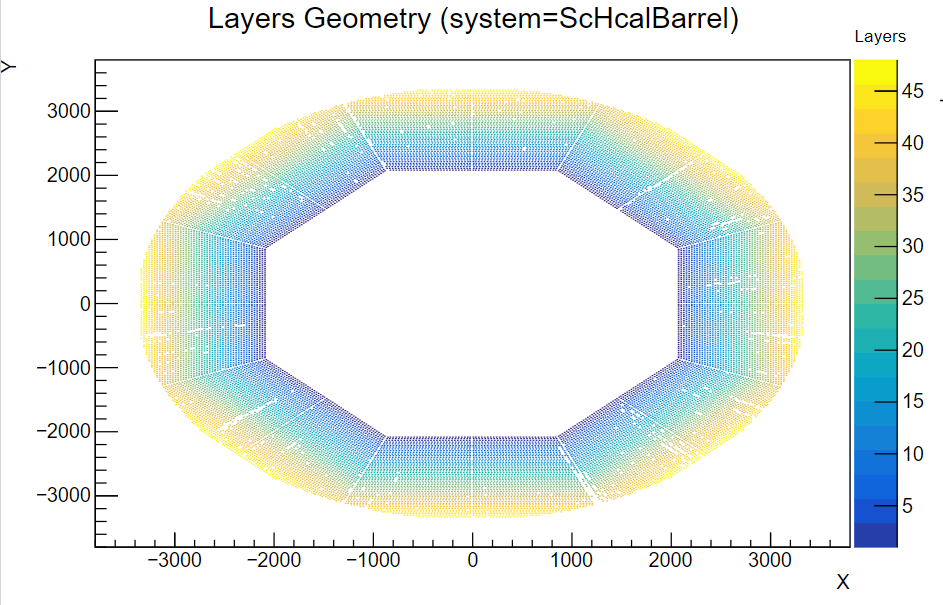}
        \caption{Layers geometry}
        \label{subfig:HCAL layers}
    \end{subfigure}%
    \caption{Layers and modules geometry of the scintillator-HCAL-Barrel system}
    \label{fig:HCAL staves-layers geometry}   
\end{figure}

In certain systems, direct selections alone are insufficient, necessitating the use of indirect methods. Taking the scintillator-HCAL-Barrel system as an illustrative example, as in the previous one, the system has a symmetry of 8 in polar angle and the radial profile can be studied through the layers. The HCAL comprises 48 layers, which, as in the previous example, are categorized into three blocks, each containing 16 layers. This is directly addressed by the direct selection on Layers, and indirectly grouping the Staves. Refer to figure \ref{fig:HCAL staves-layers geometry}.

\begin{figure}[hbtp]
    \centering
    \begin{subfigure}{0.45\textwidth}
        
        \includegraphics[width=\linewidth, height=4cm]{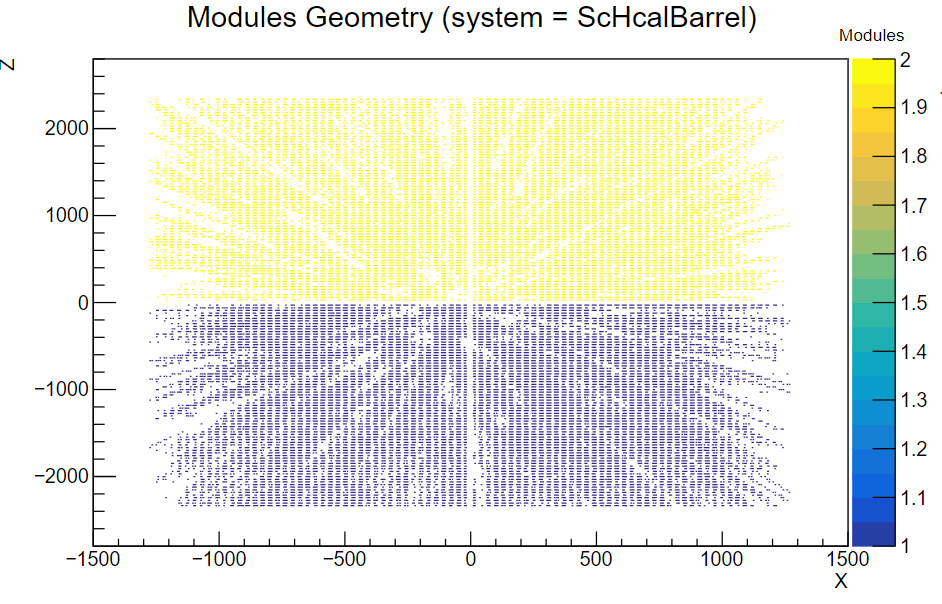}
        \caption{Modules geometry}
        \label{subfig:HCAL modules}
    
    \end{subfigure}%
    \hspace{10pt}
    \begin{subfigure}{0.45\textwidth}
        
        \includegraphics[width=\linewidth, height=4cm]{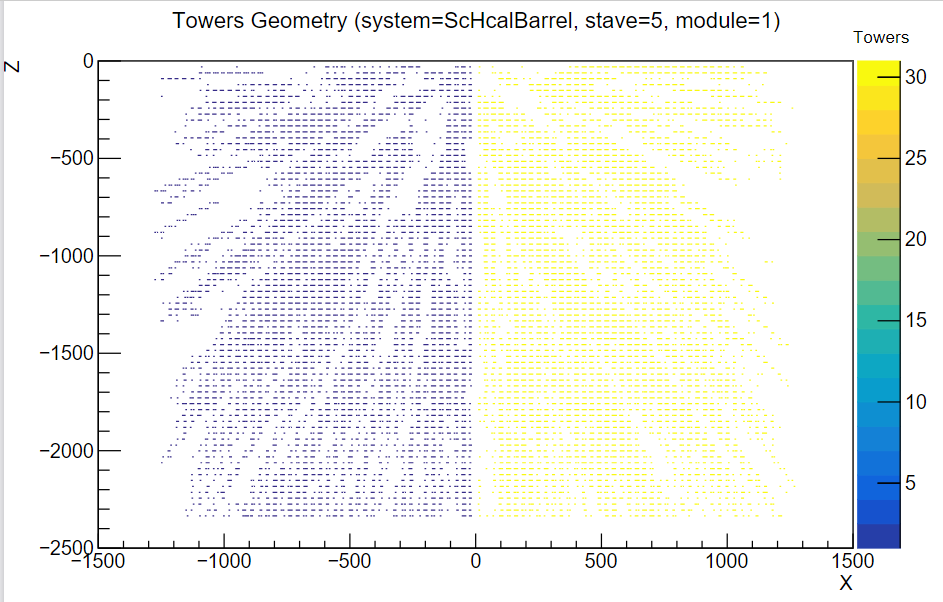}
        \caption{Towers geometry}
        \label{subfig:HCAL towers}
    
    \end{subfigure}%
    \caption{Layers and modules geometry of the scintillator-HCAL-Barrel system}
    \label{fig:HCAL modules-towers geometry}   
\end{figure}

However, for the evolution of the azimuth-angle profile, selections of modules and towers are not appropriate:  the modules indices are presented for stave 5 in figure~\ref{subfig:HCAL modules}. The HCAL barrel has only 2 modules, symmetric regarding their azimuthal angle. Each stave has 2 towers, displayed for the stave 5 in figure~\ref{subfig:HCAL towers}, symmetric in polar angle. Therefore, applying an indirect selection to cell indices becomes necessary. 
 
\begin{wrapfigure}[13]{l}{0.5\textwidth}
    \includegraphics[width=\linewidth, height=4cm]{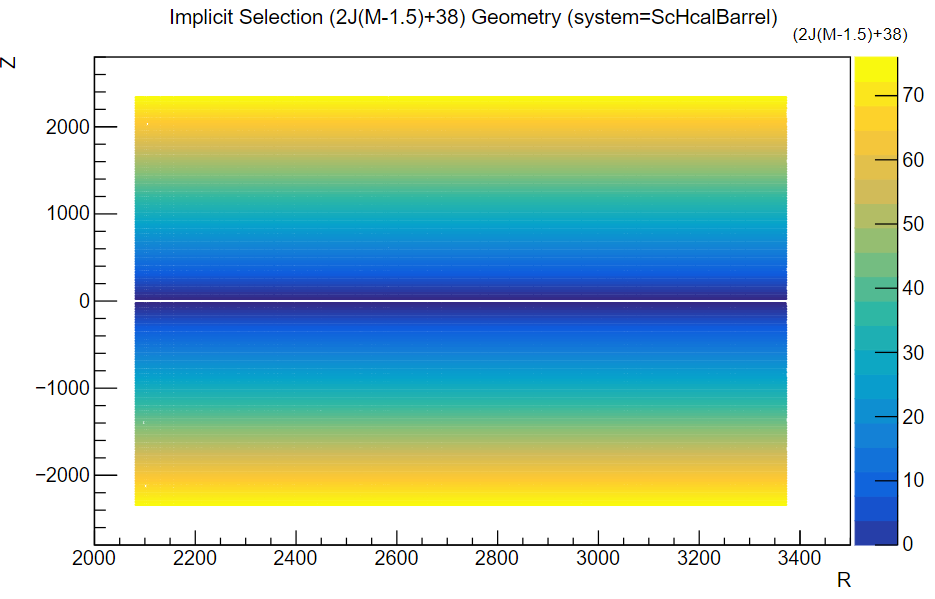}
    \caption{MJ coordinate geometry}
    \label{subfig:HCAL MJ}
\end{wrapfigure}

 A combination of M and J is defined by the equation $MJ=2J(M-1.5)+38$, where J is the logical position of the cell along $z$ and M is the module index. MJ takes values between 0 and 76, symmetric in $z$, as shown in figure~\ref{subfig:HCAL MJ}. 
 
MJ values are categorized into three groups: 0-29, 30-59, and 60-76. Thus, the total number of selections is 3+3=6 selections. From a coding perspective, a Boolean function includes hit information only if the MJ value falls within the selected block, returning 'True' in these cases. For this function, we choose the arbitrary name to be \textit{\textbf{complex\_happy}}. 

In this example, the radial profile is studied by the direct selection of three groups of layers and the polar-angle profile is looked at by an indirect selection imposed on the coordinate MJ, also by grouping them in three categories. 

\vspace{3cm}
\section{Simulated Data Samples \label{sec:data}}

The generated histograms described above count events and timed energy distributions for a given process simulated sample. 
To estimate the total event rates, hit rates or power, the histograms must be properly scaled and then summed for all (or at least the main) possible contributing physics processes or machine background. \\

For physics processes, the histogram scaling is $\sigma \times \Lint / N_\mathrm{evt}$, where $\sigma$ is the cross-section, \Lint is the instantaneous luminosity, and $N_\mathrm{evt}$ is the number of simulated events. \\

In the case of machine background, simulations are conducted over a specified number of bunch-crossings. The histogram scaling must then be $f_\mathrm{BX}/N_\mathrm{BX}$, where $f_\mathrm{BX}$ and $N_\mathrm{BX}$ are the frequency of colliding and the number of simulated bunch-crossings. \\

\begin{wrapfigure}[9]{r}{0.50\textwidth}
    \vspace{-10pt}
    \includegraphics[width=0.48\textwidth, height=4 cm]{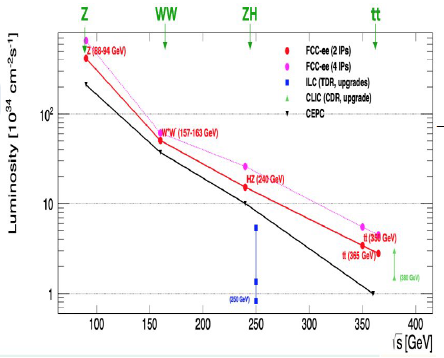}
    \caption{Luminosity of the different energy scales. At 91.2\u{GeV}, the considered run is FCC-ee (4 IPs).}
   \label{fig: luminosity}
\end{wrapfigure}

For the FCC studies, the simulated data are run at four collider configurations, defined by a center-of-mass (CM) energy, and an instantaneous luminosity, display in figure~\ref{fig: luminosity}
\begin{enumerate}
    \item \textbf{91.2 GeV:} The Z pole;
    \item \textbf{162.5 GeV:} $W^{+}W^{-}$ threshold;
    \item \textbf{240 GeV:} The resonance energy of the $ee \rightarrow ZH$ decay channel;
    \item \textbf{365 GeV:} Above the $t \Bar{t}$ threshold.
\end{enumerate}
(The details of the simulation are shown in appendix A):

\subsection{Physics Data \label{sub:physics data}}
 
For each CM energy, we simulate the dominant physics processes with the smallest biases. At 91.2\u{GeV}, the simulated Z-decay channels are (a) $ee \rightarrow qq$ and (b) $ee \rightarrow ll$ where $q = u, c, s, b, d$ and $l = \mu, \tau$.
At 162.5\u{GeV}, we simulate the channels (a) and (b) as for 91.2\u{GeV} and add the channel (c) $ee \rightarrow WW$ since we are around the resonance energy of the W production. At 240\u{GeV}, we simulate channels (a), (b), and (c) and add channel (d) $ee \rightarrow ZH$. At 365\u{GeV}, (a), (b), (c), and (d) are simulated and added to channel (e) $ee \rightarrow tt$.

\begin{wrapfigure}[13]{l}{0.5\textwidth}
    \label{fig:cross-section}
    \includegraphics[width=0.45\textwidth, height=4cm]{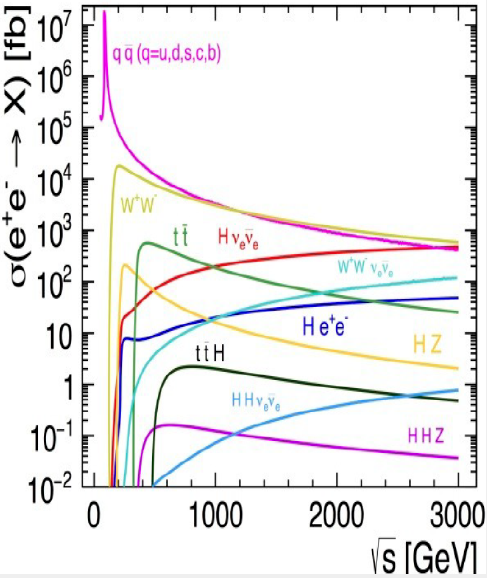}
    \caption*{Cross-section as per decay channels}
\end{wrapfigure}

The $ee \rightarrow ee$ is treated separately, due to the interference between the production and scattering processes, and the domination of the Bhabha cross-section at low angle. An angular cut of 10 degrees is applied on the resulting electron and positron to avoid the huge forward and backward scattering at low angle. The simulation phase space domain is partitioned into two parts based on the mass of the outgoing electron-positron system, set at a value of 30\u{GeV}. This partitioning allows simulating fewer events from the channel with a smaller cross-section and to save simulation time if needed. It also aids in the visualization. Since Bhabha scattering can be used in calibration and cross-check, this might be helpful. \\

The simulated data for each decay channel at each energy scale comprise 10,000 events, for which the cross-section is calculated. The data are summarized in tables (\ref{tab:91.2GeV} - \ref{tab:365GeV}).
\vspace{5pt}

\subsection{Machine Background \label{sub:machine background}}

A sizable amount of the contribution in the calorimeters can find an origin in the beam-induced backgrounds, either directly, for the parts close to the beam pipe, or indirectly via back-scattered particles on the elements at low radius.  
Several beam-induced backgrounds must be modelled~\cite{the_ild_collaboration_international_2020} (section 6.5),  \cite{behnke_international_2013} (section 1.3.1), \cite{ciarma2023beam}:
\begin{itemize}
\item Bremsstrahlung: the strong in-beam electric fields near the interaction region will produce photons and "incoherent" electron-positron pairs, mostly in the direction of the beams, but which might interact with neighbouring materials and scatter back in all the detector.  They are highly dependent on the beam bunches geometry. It is simulated using the Guinea-Pig++ package\cite{schulte1999beam}.
\item Synchrotron Radiation: arise from the beam focusing and steering in the final stages near the detector. It is usually shielded to avoid illuminating any part of the detector, but back-scattering is hard to avoid completely. It is highly dependent on the detailed implementation of elements along the beamline, and can then be estimated by e.g., the BBBrem and GuineaPig++ packages.
It was not considered in this study.
\item Muons: at linear colliders, non-Gaussian tails of the beam can interact with nearby material, creating muon pairs, which will follow the beam line all the way to the detector. They can be deflected by toroidal magnets.  
They are not considered here.
\item Neutrons: A flux of neutrons will be produced by all the above sources hitting heavy elements.  
Previous estimations show it must be low and can be neglected \cite{kleiss_bbbrem_1994}.

\end{itemize}

\begin{table}[H]

\begin{minipage}{0.5\textwidth}
    \caption{$91.2\u{GeV} \\ (N= 10000, \, \Lint = 1.4 \times 10^{-3}\u{fb^{-1} s^{-1})}$}
    \label{tab:91.2GeV}
    \begin{tabular}{ccc}
        \hline
        Channels & $\sigma$& $\sigma \times \Lint/N$ \\
        & ($10^{5} \u{fb}$)& $(\u{s^{-1}})$ \\
        \hline
        $ee \rightarrow qq$ & 344 & 4.82 \\
        $ee \rightarrow ll$ & 34.6 & 0.484 \\
        $ee \rightarrow ee$ \\ $(M_{ee} < 30\u{GeV})$ & 1.01 & 0.0141  \\
        $ee \rightarrow ee$ \\ $(M_{ee} > 30\u{GeV})$ & 57.8 & 0.809 \\
        \hline
    \end{tabular}
\end{minipage}\hfill
\hspace{10pt}
\begin{minipage}{0.5\textwidth}
    \vspace{11.5pt}
    \caption{$162.5\u{GeV} \\ (N= 10000, \, \Lint = 2.14 \times 10^{-4} \u{fb^{-1} s^{-1}})$}
    \label{tab:162.5GeV}
    \begin{tabular}{ccc}
        \hline
        Channels & $\sigma$ & $\sigma \times \Lint/ N$ \\
        & ($10^{5}\u{fb}$) & $(\u{s^{-1}})$ \\
        \hline
        $ee \rightarrow qq$ & 1.55 & $3.32 \times 10^{-3}$ \\
        $ee \rightarrow ll$ & 0.241 & $5.16 \times 10^{-4}$ \\
        $ee \rightarrow WW$ & 0.0504 & $1.08 \times 10^{-4}$ \\
        $ee \rightarrow ee$ \\ $(M_{ee} < 30 GeV)$ & 0.240 & $5.14 \times 10^{-4}$ \\
        $ee \rightarrow ee$ \\ $(M_{ee} > 30 GeV)$ & 12.9 & $2.76 \times 10^{-2}$\\
        \hline
    \end{tabular}
\end{minipage}

\begin{minipage}{0.5\textwidth}
    \caption{$240\u{GeV} \\ (N= 10000, \, \Lint = 6.9 \times 10^{-5} \, \u{fb^{-1}} \, \u{s^{-1}})$}
    \label{tab:240GeV}
    \begin{tabular}{ccc}
        \hline
        Channels & $\sigma$ & $\sigma \times \Lint/N$ \\
        & ($ 10^{5} \, \u{fb}$) & $(\u{s^{-1}})$ \\
        \hline
        $ee \rightarrow qq$ & 0.550 & $3.80 \times 10^{-4}$ \\
        $ee \rightarrow ll$ & 0.100 & $6.88 \times 10^{-5}$ \\
        $ee \rightarrow WW$ & 0.167 & $1.15 \times 10^{-4}$ \\
        $ee \rightarrow ZH$ & 0.00204 & $1.41 \times 10^{-6}$ \\
        $ee \rightarrow ee$ \\ $(M_{ee} < 30\u{GeV})$ & 0.120 & $8.29 \times 10^{-5}$ \\
        $ee \rightarrow ee$ \\ $(M_{ee} > 30\u{GeV})$ & 5.92 & $4.09 \times 10^{-3}$ \\
        \hline
    \end{tabular}
\end{minipage}\hfill
\hspace{10pt}
\begin{minipage}{0.5\textwidth}
    \vspace{11.5pt}
    \caption{$365\u{GeV} \\ (N= 10000, \: \Lint = 1.2 \times 10^{-5}\u{fb^{-1} s^{-1}})$}
    \label{tab:365GeV}
    \begin{tabular}{ccc}
        \hline
        Channels & $\sigma$& $\sigma \times \Lint/N$ \\
        & ($10^{5}\u{fb}$)& $(\u{s^{-1}})$ \\
        \hline
        $ee \rightarrow qq$ & 0.228 & $2.74 \times 10^{-5}$ \\
        $ee \rightarrow ll$ & 0.0430 & $5.16 \times 10^{-6}$ \\
        $ee \rightarrow WW$ & 0.111 & $1.33 \times 10^{-5}$ \\
        $ee \rightarrow ZH$ & 0.00123 &  $1.47 \times 10^{-7}$ \\
        $ee \rightarrow tt$ & 0.00372 & $4.46 \times 10^{-7}$ \\
        $ee \rightarrow ee$ \\ $(M_{ee} < 30\u{GeV})$ & 0.0499 & $5.99 \times 10^{-6}$ \\
        $ee \rightarrow ee$ \\ $(M_{ee} > 30\u{GeV})$ & 2.58 & $4.65 \times 10^{-6}$ \\
        \hline
    \end{tabular}
\end{minipage}

\end{table}

\clearpage 
\newpage
\section{Results \label{sec:Results}}
 
The histograms are obtained for 11 different systems with various direct and indirect selections; just a few of them are displayed here as an illustration\footnote{The full set of histograms can be found here: \url{https://github.com/LLR-ILD/CalorimeterFluxes/tree/main/Histograms}}. The plots shown in this section correspond to all the merged physics processes at a specific energy, each scaled by its weight according to tables (\ref{tab:91.2GeV}-\ref{tab:365GeV}) along with the weighted machine background. 
\subsection{\label{sub:Angular Distribution}Angular Distribution}
The number-of-hits histograms for the ECAL Barrel at angle $\theta = 90^{\circ}$ (module 3) show that most of the hits are in the first 2 layer blocks (the first 20 layers).  
\begin{figure}[H]
    \centering
    \begin{subfigure}{0.45\textwidth}
        
        \includegraphics[width=\linewidth, height=6cm]{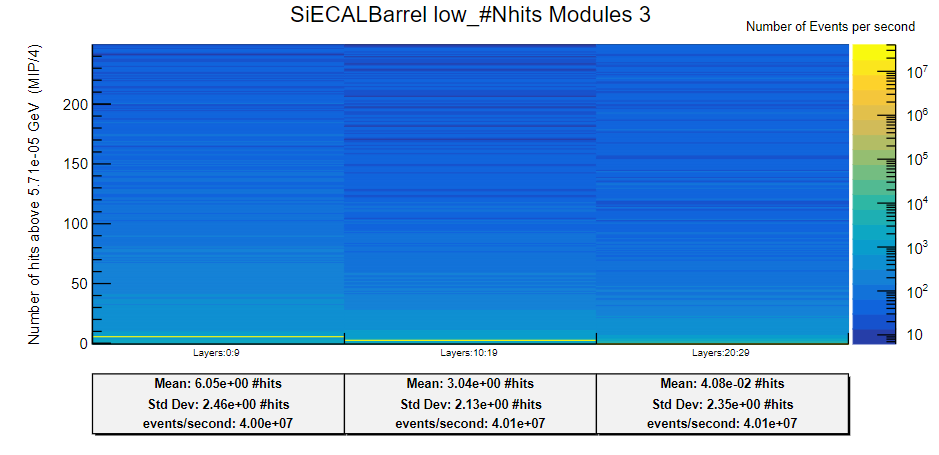}
        \caption{Lower-scale}
        \label{subfig:low-scale silicon}
    
    \end{subfigure}%
    \hspace{10pt}
    \begin{subfigure}{0.45\textwidth}
        
        \includegraphics[width=\linewidth, height=6cm]{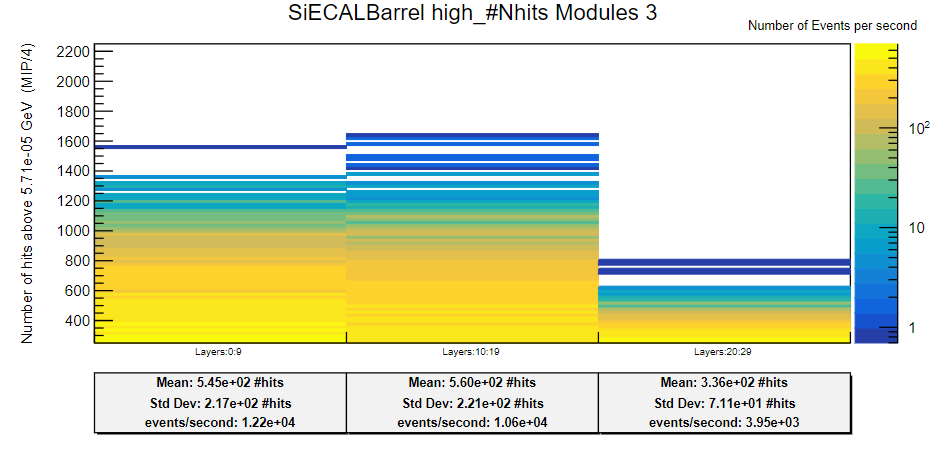}
        \caption{Upper-scale}
        \label{subfig:upper-scale silicon}
    
    \end{subfigure}%
    \caption{Lower and upper scales of the number-of-hits-above-(MIP/4) histograms for 3 different layers blocks in module 3 of the silicon-ECAL-Barrel system at $\sqrt{s} = 91.2 \text{ GeV}$}
    \label{fig:layers silicon ECAL }   
\end{figure}

To study the polar-angular behaviour, the same histogram types are shown but for the first layer block (It is one of the blocks with most of the hits) with all the modules as further selections.

\begin{figure}[H]
    \centering
    \begin{subfigure}{0.45\textwidth}
        
        \includegraphics[width=\linewidth, height=6cm]{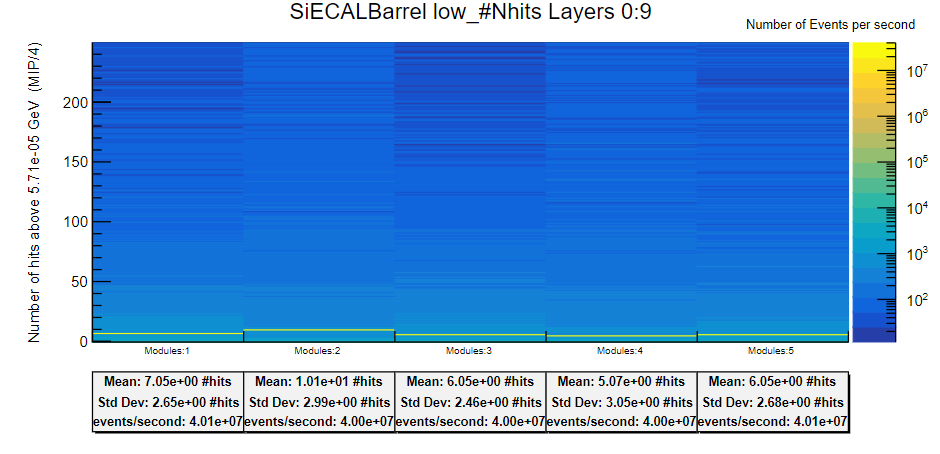}
        \caption{Lower-scale}
        \label{subfig:low-scale silicon L1}
    
    \end{subfigure}%
    \hspace{10pt}
    \begin{subfigure}{0.45\textwidth}
        
        \includegraphics[width=\linewidth, height=6cm]{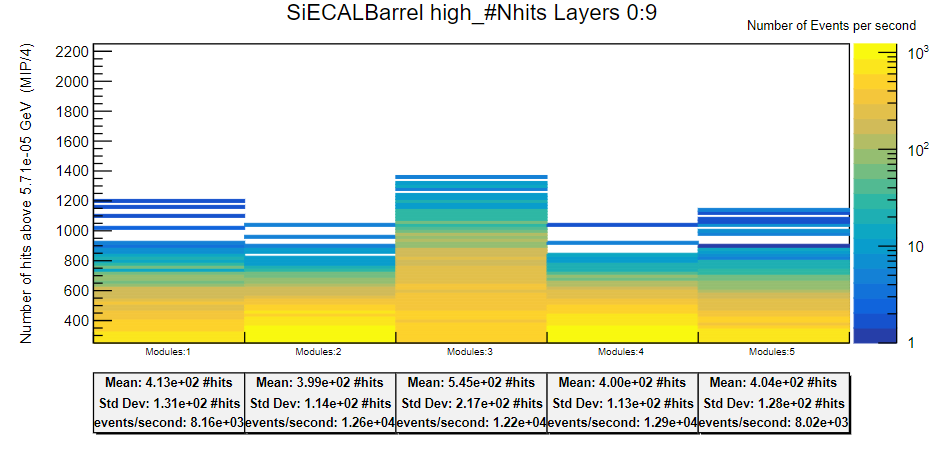}
        \caption{Upper-scale}
        \label{subfig:upper-scale silicon L1}
    
    \end{subfigure}%
    \caption{Lower and upper scales of the number-of-hits-above-(MIP/4) histograms for 5 different modules in the first 10-layer block of the silicon-ECAL-Barrel system at $\sqrt{s} = 91.2 \text{ GeV}$}
    \label{fig:modules silicon ECAL}   
\end{figure}
The conclusion is that there is no significant angular dependence as all the modules exhibit similar statistics except for the central module. The uniqueness in the central-module behaviour traces back to the fact that at 91.2\u{GeV}, the output particles are produced back-to-back, giving more hits near $\theta = 90^{\circ}$. \\ 

\begin{figure}[H]
    \centering
    \begin{subfigure}{0.45\textwidth}
        
        \includegraphics[width=\linewidth, height=5cm]{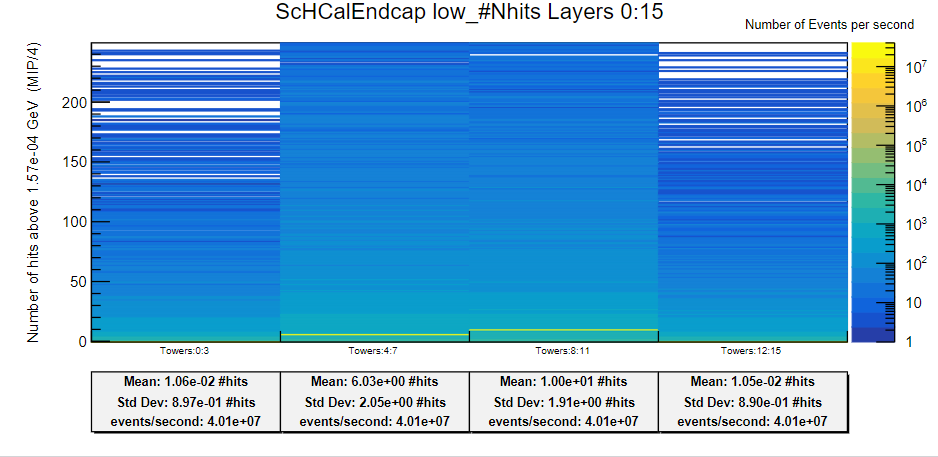}
        \caption{Lower-scale}
        \label{subfig:low-scale scintillator L1}
    
    \end{subfigure}%
    \hspace{10pt}
    \begin{subfigure}{0.45\textwidth}
        
        \includegraphics[width=\linewidth, height=5cm]{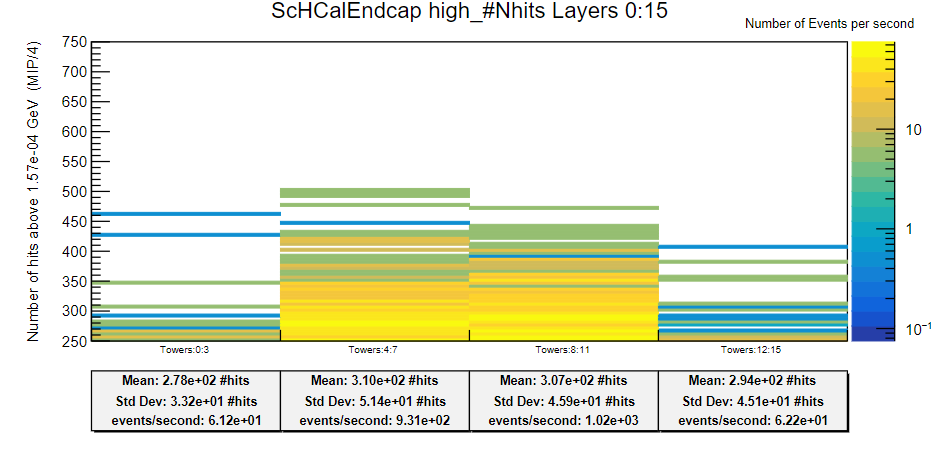}
        \caption{Upper-scale}
        \label{subfig:upper-scale scintillator L1}
    
    \end{subfigure}%
    \caption{Lower and upper scales of the number-of-hits-above-(MIP/4) histograms for 4 different towers selections in the first 16-layer block of the scintillator-HCAL-Endcap system at $\sqrt{s} = 91.2 \text{ GeV}$}
    \label{fig:towers scintillator HCAL}   
\end{figure}

In contrast, the analysis of the HCAL Endcap leads to different conclusions. First, like the ECAL, most of the hits are not recorded in the third layers block (the last 16 layers). This can be seen from the central towers (towers 4-7 are chosen). Central towers are the closest to the beam pipe, and hence they get most of the hits.  
\begin{figure}[H]
    \centering
    \begin{subfigure}{0.45\textwidth}
        
        \includegraphics[width=\linewidth, height=6cm]{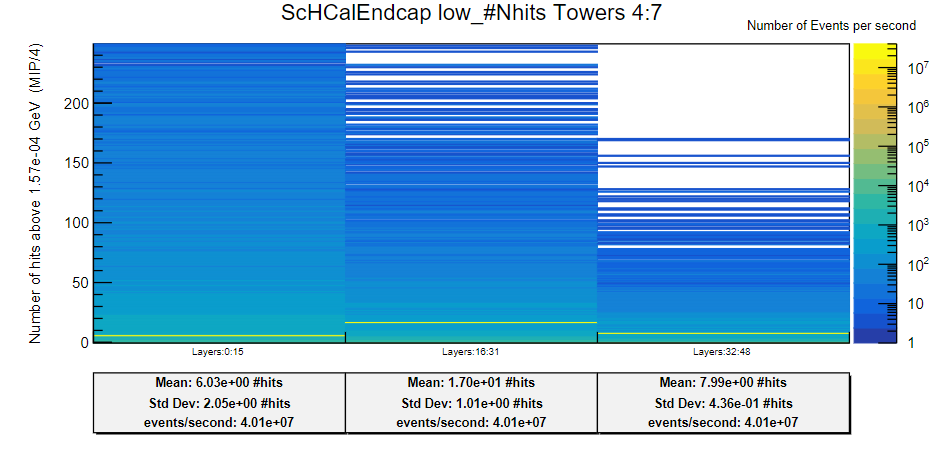}
        \caption{Lower-scale}
        \label{subfig:low-scale scintillator}
    
    \end{subfigure}%
    \hspace{10pt}
    \begin{subfigure}{0.45\textwidth}
        
        \includegraphics[width=\linewidth, height=6cm]{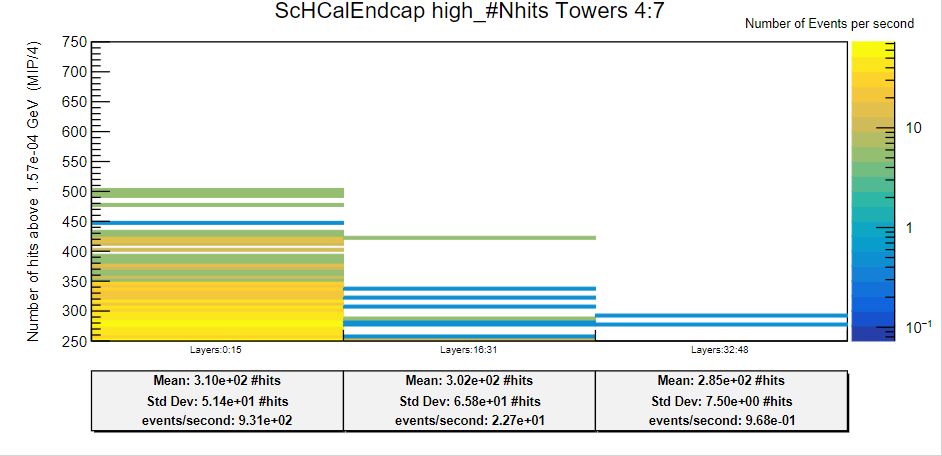}
        \caption{Upper-scale}
        \label{subfig:upper-scale scintillator}
    
    \end{subfigure}%
    \caption{Lower and upper scales of the number-of-hits-above-(MIP/4) histograms for 3 different layers blocks in towers 4-7 of the scintillator-HCAL-Endcap system at $\sqrt{s} = 91.2 \text{ GeV}$}
    \label{fig:layers scintillator HCAL}   
\end{figure}
Referring to Figure \ref{fig:towers scintillator HCAL}, the right side is roughly symmetrical to the left side in the lower and upper scale plots, consistent with the prediction that the line of symmetry here originates from the beam pipe. However, if our analysis is limited to one side only (i.e., the first 8 towers or the last 8 towers), it becomes evident that the tower selections are asymmetric, and the angular dependency is significant here, unlike in the case of the ECAL Barrel, where all modules exhibited symmetric distributions except for the central module, as discussed. This difference is attributed to the fact that the central towers (i.e., towers 4-7 or towers 8-11) are closer to the beam pipe than the other towers, and hence, they receive more hits.
\subsection{\label{sub:Dynamic Range}Dynamic Range}
Another important result that can be extrapolated from these plots is the dynamic range represented by the maximum energy in the unit of MIP. The questions that need to be answered are what is the maximum energy to be covered by the electronics and if it also depends on the beam energy. Thus, the histogram type that is needed here is the scaled upper-scale energy. \\ 

\begin{figure}[H]
    \centering
    \begin{subfigure}{0.45\textwidth}
        
        \includegraphics[width=\linewidth, height=6cm]{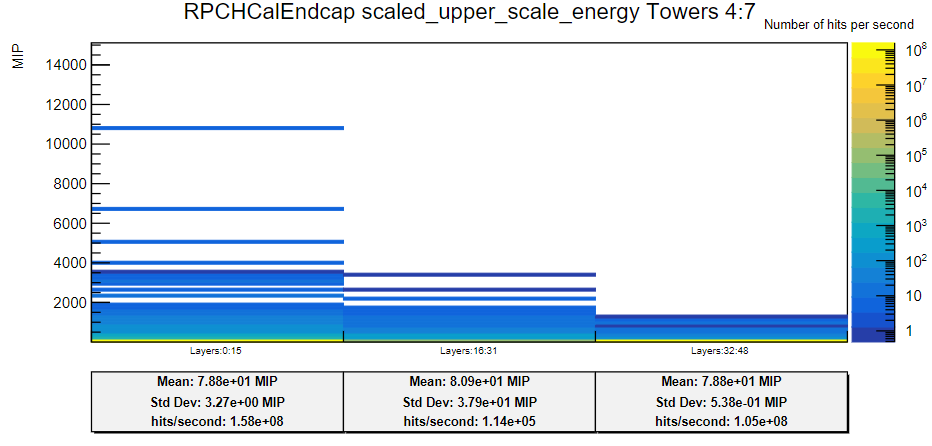}
        \caption{$\sqrt{s} = 91.2 \text{ GeV}$}
        \label{subfig:dynamic range RPC 91.2 GeV}
    
    \end{subfigure}%
    \hspace{10pt}
    \begin{subfigure}{0.45\textwidth}
        
        \includegraphics[width=\linewidth, height=6cm]{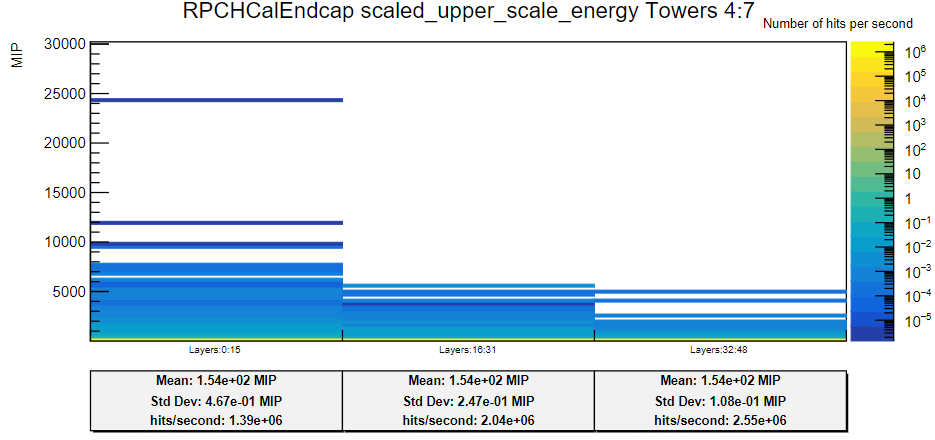}
        \caption{$\sqrt{s} = 240\u{GeV}$}
        \label{subfig:dynamic range RPC 240 GeV}
    
    \end{subfigure}%
    \caption{Scaled upper-scale energy distributions in the MIP unit of 3 different layer blocks at towers 4-7 of the RPC-HCAL-Endcap system at $\sqrt{s} = 91.2\u{GeV}$ and $\sqrt{s}=240\u{GeV}$}
    \label{fig:dynamic range RPC HCAL endcap}   
\end{figure}

It can be seen that the maximum energy falls in the same range for both $\sqrt{s} = 91.2 \text{ GeV}$ and $\sqrt{s}=240 \text{ GeV}$. The example chosen is tower 0 ($\theta = 90^{\circ}$) in the silicon-ECAL-Endcap system; however, different tower selections would yield similar results. 
\begin{figure}[H]
    \centering
    \begin{subfigure}{0.45\textwidth}
        
        \includegraphics[width=\linewidth, height=6cm]{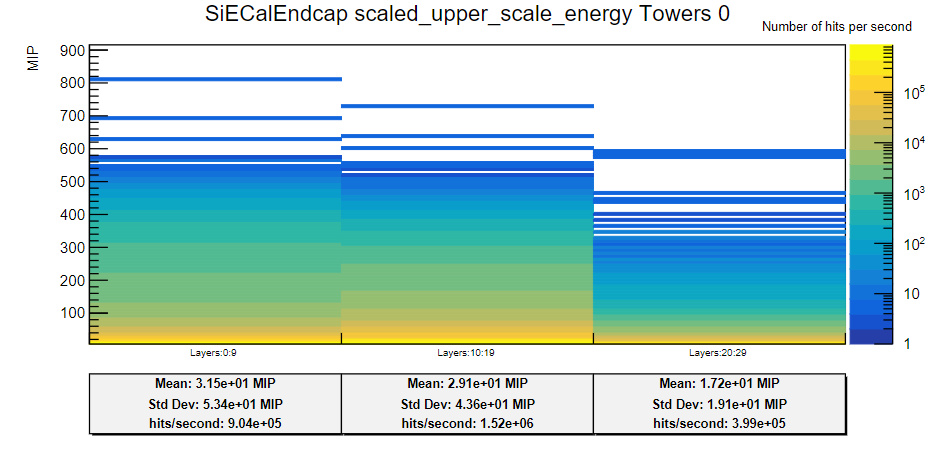}
        \caption{$\sqrt{s} = 91.2 \text{ GeV}$}
        \label{subfig:dynamic range 91.2 GeV}
    
    \end{subfigure}%
    \hspace{10pt}
    \begin{subfigure}{0.45\textwidth}
        
        \includegraphics[width=\linewidth, height=6cm]{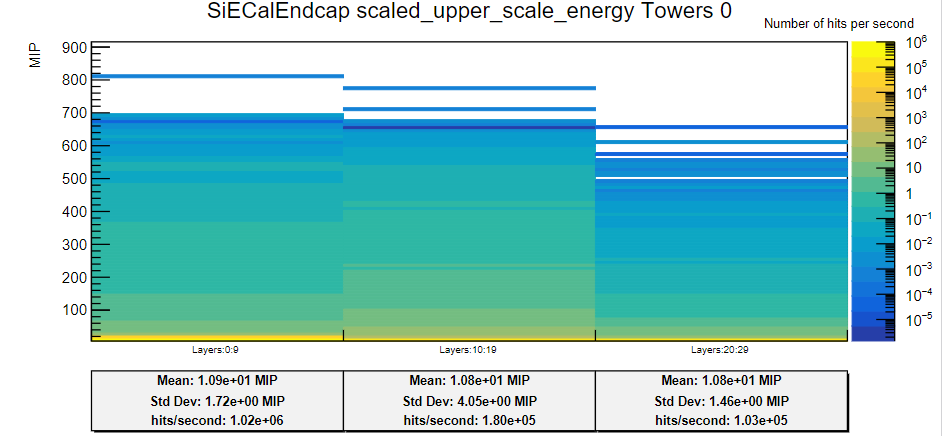}
        \caption{$\sqrt{s} = 240 \text{ GeV}$}
        \label{subfig:dynamic range 240 GeV}
    
    \end{subfigure}%
    \caption{Scaled upper-scale energy distributions in the MIP unit of 3 different layer blocks at tower 0 of the silicon-ECAL-Endcap system at $\sqrt{s} = 91.2 \text{ GeV}$ and $\sqrt{s}=240 \text{ GeV}$}
    \label{fig:dynamic range silicon ecal endcap}   
\end{figure}
This behaviour is not obeyed in the RPC. Figure \ref{fig:dynamic range RPC HCAL endcap} shows 2 things: 
\begin{enumerate}
    \item The energy range is much bigger for the RPC compared to the silicon. The maximum energy was around $800\u{MIP}$ for the silicon, whereas for the RPC system, it reached $11000\u{MIP}$ and $25000\u{MIP}$ for beam energy of 91.2 GeV and 240 GeV, respectively.
    \item The beam energy makes much difference in the dynamic range for the RPC. 
\end{enumerate}

\newpage
\section*{Conclusion}
In this project, a versatile software package was introduced to produce different sets of histograms of different parts of the calorimeter. At four different energy scales and different runs, dominating physics processes and machine background were simulated in ILD to capture the full output of the calorimeter distributions. Different results were shown of applying the package on the output physics. \\ \\
The future work will entail an extension of this package to other parts of the ILD detector, such as the tracker. Additionally, an expansion can be applied on the calorimeters of other detectors rather than the ILD. Finally, on the technical side, LCIO \cite{LCIO} package is to be replaced with EDM4hep \cite{EDM4hep} which is the common event-data model of key4HEP stack software \cite{key4hep} aiming for more coherence with other HEP detectors and experiments. 
\newpage
\appendix
\section*{\label{appendix section: Physics Data Simulation}Appendix A: Physics Data Simulation}
At the generator level, the data is simulated by Whizard 3.0.3 \cite{whizard}. The hadronization is simulated by Pythia6 \cite{pythia}. Whizard and Pythia are integrated together using EventProducer package \cite{EventProducer}. \\ \\
At the detector level, iLCSoft software framework \cite{iLCSoft} is used. The used version is v02-02-02. DD4hep \cite{DD4hep} is the common detector geometry description in iLCSoft which is component based. DDG4 \cite{DDG4} is the component affiliated with the full simulation, which relies on Geant4 \cite{Genat4}. Thus, the ILD geometry is described using DD4hep as explained above. Afterward, a python program called ddsim, being a part of DD4hep, is used to run the full simulation. The detector geometry has multiple versions. The one used here is ILD\_l5\_v02.\\ \\
The output of the full simulation is given in LCIO format, which is the common event-data model of iLCSoft. Our software package starts operating by decoding the slcio files (LCIO files format) and retrieving the information to be fed into the selected histograms.
\newpage

\bibliographystyle{IEEEtran}
\bibliography{sorsamp}
\end{document}